\documentstyle[aps,prbbib,eclepsf]{revtex}

\begin{document}
\draft

\title{ Theory of tunneling spectroscopy in superconducting 
Sr$_{2}$RuO$_{4}$  }

\author{Masashi Yamashiro and Yukio Tanaka}

\address{Graduate School of Science and Technology,  
Niigata University, Ikarashi, 
Niigata 950-21, Japan }

\author{S. Kashiwaya}
\address{Electrotechnical Laboratory, Tsukuba, Ibaraki 305, Japan}
\date{\today}
\maketitle
\begin{abstract}
A theory for 
tunneling spectroscopy in normal metal
/insulator/triplet superconductor junction 
is presented. 
We assume two kinds of non-unitary triplet superconducting states
which are the most promising states for Sr$_{2}$RuO$_{4}$. 
The calculated conductance spectra show
zero-bias peaks as well as gap structures.
The existences of residual components in the spectra 
reflect the non-unitary properties of superconducting states.
\end{abstract}
\vspace{24pt}

\widetext

Recent discovery of superconductivity in Sr$_{2}$RuO$_{4}$\cite{maeno} 
provides us the first example of a noncuprate layered perovskite 
material that exhibits superconductivity. Since this compound 
is isostructural to the cuprate superconductors, the electronic 
properties in the normal state \cite{maeno2}
and superconducting state\cite{yoshida1} are highly anisotropic. 
The rather large residual density of states 
of quasiparticles at low temperatures is indicated by several experiments.
\cite{maeno3,ishida} Furthermore, there are several 
evidences which support the indications of 
ferromagnetic spin fluctuations.\cite{rice} 
Based on these facts, 
some theories\cite{sigrist,machida} proposed that 
the non-unitary triplet pairing superconducting states 
are realized in Sr$_{2}$RuO$_{4}$. 
Since the triplet pairing states have strong anisotropy in $k$-space, 
novel interference effects of the quasiparticles
are expected to occur at boundaries and surfaces. 
To determine the symmetry of the 
pair potential definitively, it is important to predict the spectra of 
tunneling experiments which play a significant role 
to identify the $d$-wave symmetry in 
the high-$T_{C}$ superconductors.
\cite{tanaka1,kashiwaya1,kashiwaya2} \par
Recently a tunneling conductance formula  
for normal metal/insulator/anisotropic singlet 
superconductor junctions 
was presented.\cite{tanaka1,kashiwaya2} 
Even in the case of an spin-singlet superconductor, 
when the pair potential becomes 
anisotropic \cite{bruder} 
and changes its sign on the Fermi surface, 
zero-energy states \cite{hu} 
appear at the surface
depending on the orientation of the surface. 
The formation of the zero-energy states \cite{hu} 
induces zero-bias conductance peaks in
tunneling spectroscopy, which were actually observed in the 
experiments of high-$T_{C}$ superconductors.\cite{kashiwaya1,geerk} 
By assuming  $d_{x^{2}-y^{2}}$-wave symmetry of 
the pair potentials, 
not only the zero-bias conductance peaks but also gap like spectra 
were systematically explained.\cite{tanaka1,kashiwaya1,kashiwaya2} 
However, the 
tunneling conductance for 
normal metal / insulator /triplet superconductor($N/I/TS$) 
junction is not well clarified yet. \par
In the present paper, 
we present a formulation of the tunneling conductance spectra 
of $N/I/TS$ junction
by extending the previous one
for anisotropic singlet 
superconductors.\cite{tanaka1,kashiwaya1,kashiwaya2} 
Although the superconducting states of 
Sr$_{2}$RuO$_{4}$ are not clarified yet, 
we will choose 
two kinds of triplet $p$-wave pair potentials 
($E_{u}$ states) which are proposed by Machida $et$. $al$. \cite{machida}
and Sigrist and Zhitomirsky.\cite{sigrist} 
Large variety of conductance spectra including zero-bias conductance peaks 
are obtained depending on the tunneling directions.
Thus, the tunneling spectroscopy measurements is one of 
the useful 
method to identify the pairing symmetry of Sr$_{2}$RuO$_{4}$. \par
For the calculation, we assume a 
$N/I/TS$ junction model in
the clean limit with semi-infinite double layer structure. 
We also assume a nearly two-dimensional Fermi momentum  
by restricting the $z$ component of the Fermi surface to the 
region given by $-\bar{\delta}< \sin^{-1}(k_{Fz}/k_{F})<\bar{\delta}$. 
The flat interface is perpendicular to the $x$-axis, 
and is located at $x=0$ (Fig.1A). 
The barrier potential at the interface has a delta-functional form 
$H\delta(x)$, where $\delta(x)$ and $H$ are 
the delta-function and its amplitude, respectively.
Similarly, we consider alternative situation that the flat interface 
is perpendicular to the $z$-axis and is located at $z=0$ (Fig.1B).
The Fermi wave number 
$k_{F}$ and the effective mass $m$ are assumed to be equal both in 
the normal metal and in the superconductor. The wave function of 
the quasiparticles in inhomogeneous anisotropic superconductors is 
given by the solution of 
the Bogoliubov-de Gennes(BdG) equation.\cite{bruder,hu} 
Although this equation includes a non-local pair potential with two position 
coordinates for the Cooper pairs, we assume that the effective 
pair potential is given by 
\begin{equation}
\Delta_{\rho\rho^{\prime}}(\mbox{\boldmath$k,r$})=\left\{ \begin{array}{ll}
\Delta_{\rho\rho^{\prime}}(\theta, \phi)\Theta(x), \hspace{12pt}
\mbox{$z$-$y$ plane interface}\\[10pt]
\Delta_{\rho\rho^{\prime}}(\theta, \phi)\Theta(z), \hspace{12pt}
\mbox{$x$-$y$ plane interface}
\end{array}, \right. \hspace{12pt}
\frac{k_{x}+ik_{y}}{\mid \mbox{\boldmath$k$} \mid}=\sin\theta e^{i\phi}, 
\hspace{12pt} \frac{k_{z}}{\mid \mbox{\boldmath$k$} \mid}=\cos\theta
\label{eqn:e1}
\end{equation} 

\noindent
where $\theta$ is the polar angle and $\phi$ is the azimuthal angle 
in the $x$-$y$ plane. The quantities $\rho$ and $\rho^{\prime}$ denote spin indices. 
This pair potential is rather simplified by applying the quasi-classical approximation
and by ignoring the pair breaking effect at the interface.
\cite{bruder}
In Eq.(\ref{eqn:e1}), 
$\mbox{\boldmath{$k$}}$ is a wave vector of the relative motion of the 
Cooper pairs and is fixed on the Fermi surface($|\mbox{\boldmath{$k$}}|$=
$k_{F}$). 
The quantities $\Theta(x)$, $\Theta(z)$ and $\mbox{\boldmath{$r$}}$ 
are the Heviside step functions 
and the center of mass coordinate of 
the pair potentials, respectively. \par
Suppose an electron is injected from the normal metal with angles 
$\theta$ and $\phi$. 
We have taken care of the fact that the momentum parallel to the 
interface is conserved at the interface. 
The electron injected from the normal metal is reflected as an electron 
(normal reflection) and a hole (Andreev reflection). 
When the interface 
is perpendicular to the $x$-axis($z$-$y$ plane interface) (Fig.1A), 
the transmitted holelike quasiparticle(HLQ) and 
electronlike quasiparticle(ELQ) feel different effective 
pair potentials $\Delta_{\rho\rho'}(\theta, \phi_{+})$ 
and $\Delta_{\rho\rho'}(\theta, \phi_{-})$, 
with $\phi_{+}=\phi$ and $\phi_{-}=\pi-\phi$. 
On the other hand, 
in the case when the interface is perpendicular 
to the  $z$-axis($x$-$y$ plane interface), 
two kinds of quasiparticles feel 
$\Delta_{\rho\rho'}(\theta_{+}, \phi)$ 
and $\Delta_{\rho\rho'}(\theta_{-}, \phi)$, 
with $\theta_{+}=\theta$ and $\theta_{-}=\pi-\theta$, 
(Fig.1B) respectively. 
The coefficients of the Andreev reflection 
$a_{\rho\rho'}(E,\theta,\phi)$
and normal reflection 
$b_{\rho\rho'}(E,\theta,\phi)$ 
are determined 
by solving the BdG equations under 
the following boundary conditions
\begin{equation}
\begin{array}{c}
\left. \Psi(\mbox{\boldmath$r$})\right|_{x=0_{-}}=
\left. \Psi(\mbox{\boldmath$r$})\right|_{x=0_{+}},\hspace{20pt}
\displaystyle
\left. \frac{d\Psi(\mbox{\boldmath$r$})}{dx}\right|_{x=0_{-}}=
\left. \frac{d\Psi(\mbox{\boldmath$r$})}{dx}\right|_{x=0_{+}}-
\left. \frac{2mH}{\hbar^{2}}\Psi(\mbox{\boldmath$r$})\right|_{x=0_{-}}
\end{array}
\label{eqn:e2}
\end{equation}
for $z$-$y$ plane interface and 
\begin{equation}
\begin{array}{c}
\left. \Psi(\mbox{\boldmath$r$})\right|_{z=0_{-}}=
\left. \Psi(\mbox{\boldmath$r$})\right|_{z=0_{+}},\hspace{20pt}
\displaystyle
\left. \frac{d\Psi(\mbox{\boldmath$r$})}{dz}\right|_{z=0_{-}}=
\left. \frac{d\Psi(\mbox{\boldmath$r$})}{dz}\right|_{z=0_{+}}-
\left. \frac{2mH}{\hbar^{2}}\Psi(\mbox{\boldmath$r$})\right|_{z=0_{-}}
\end{array}
\label{eqn:e3}
\end{equation}
for $x$-$y$ plane interface. 
Using the obtained coefficients, the normalized tunneling conductance 
is calculated according to the formula given by our previous works
\cite{tanaka1,kashiwaya2}
\begin{equation}
\sigma(E)=\left\{ \begin{array}{ll}
\displaystyle
\frac{\int_{\pi/2-\bar{\delta}}^{\pi/2}\!\int_{-\pi/2}^{\pi/2}
(\sigma_{S,\uparrow} + \sigma_{S,\downarrow})\sigma_{N}
\sin^{2}\theta 
\cos\phi d\theta d\phi}{\int_{\pi/2-\bar{\delta}}^{\pi/2}
\!\int_{-\pi/2}^{\pi/2}2\sigma_{N}
\sin^{2}\theta \cos\phi d\theta d\phi} & \hspace{12pt} 
\mbox{$z$-$y$ plane 
interface}\\[10pt]
\displaystyle
\frac{\int_{\pi/2-\bar{\delta}}^{\pi/2}\!\int_0^{2\pi}
(\sigma_{S,\uparrow} + \sigma_{S,\downarrow})
\sigma_{N}\sin\theta \cos\theta d\theta d\phi}
{\int_{\pi/2-\bar{\delta}}^{\pi/2}\!\int_0^{2\pi}2\sigma_{N}\sin\theta 
\cos\theta d\theta d\phi} & 
\hspace{12pt} \mbox{$x$-$y$ plane interface}
\end{array} \right.
\label{eqn:e4}
\end{equation}
where $\sigma_{N}$ denotes the normal state tunneling conductance given by
\begin{equation}
\sigma_{N}=\left\{ \begin{array}{ll}
\displaystyle
\frac{\sin^{2}\theta\cos^{2}\phi}{\sin^{2}\theta\cos^{2}\phi+Z^{2}} 
& \hspace{12pt} \mbox{ $z$-$y$ plane interface}\\[10pt]
\displaystyle
\frac{\cos^{2}\theta}{\cos^{2}\theta+Z^{2}} & \hspace{12pt}
\mbox{$x$-$y$ plane interface}\end{array} ,\right. 
\hspace{24pt}Z=\frac{mH}{\hbar^{2}k_{F}}.
\label{eqn:e5}
\end{equation}
In the above, 
$E$ denotes an energy of quasi-particles measured from Fermi energy. 
The quantity $\sigma_{S,\rho}$ is given as 
\begin{equation}
\sigma_{S,\rho}=\frac{
1 + \mid a_{\uparrow\rho} \mid^{2} + 
\mid a_{\downarrow\rho} \mid^{2} 
- \mid b_{\uparrow\rho} \mid^{2} - 
\mid b_{\downarrow\rho} \mid^{2}}{\sigma_{N}}. 
\label{eqn:e6}
\end{equation}
Hereafter, following  the discussions by Sigrist \cite{sigrist} 
and Machida, \cite{machida} 
we will choose two kinds of non-unitary pair potentials 
with tetragonal symmetry. 
These two types of $E_{u}$ symmetry states are independent of 
$k_{z}$
due to the two-dimensional nature of the Fermi surface. 
Both of these have a matrix form of the pair potential, 
with 
$\Delta_{\uparrow\uparrow}(\theta,\phi)=\Lambda_{i}(\theta, \phi)$, 
$\Delta_{\uparrow\downarrow}(\theta,\phi)=
\Delta_{\downarrow\uparrow}(\theta,\phi)=
\Delta_{\downarrow\downarrow}(\theta,\phi)=0$, 
where $\Lambda_{i}(\theta, \phi)$ is the 
orbital part which is reduced to 
depend on $\theta$ and $\phi$. 
Two kinds of $\Lambda_{i}(\theta, \phi)$ are given by 
$\Lambda_{1}(\theta, \phi)=\Delta_{0}\sin\theta(\sin\phi+\cos\phi)$ 
and $\Lambda_{2}(\theta, \phi)=\Delta_{0}\sin\theta e^{i\phi}$, 
where $\Delta_{0}$ is the absolute value of the pair potential 
in a bulk superconductor. 
For the abbreviation, 
we will call the superconducting state  of the pair potential 
with $\Lambda_{1}(\theta,\phi)$ 
($\Lambda_{2}(\theta,\phi)$) 
as $E_{u}$(1) ($E_{u}$(2)) state in the following. 
Normalized conductance $\sigma_{S,\uparrow}$ is described as follows, 
\noindent
\begin{itemize}
\item $E_{u}$(1)
\begin{eqnarray}
\displaystyle
\sigma_{S,\uparrow}
\displaystyle
&=&\frac{1+\sigma_{N}|\Gamma_{+}|^{2}+
(\sigma_{N}-1)|\Gamma_{+}|^{2}|\Gamma_{-}|^{2}}
{|1+(\sigma_{N}-1)\Gamma_{+}\Gamma_{-}|^{2}} 
 \hspace{12pt} \mbox{$z$-$y$ plane interface}
\label{eqn:e7} \\[12pt]
\displaystyle
&=&\frac{1+\sigma_{N}|\Gamma_{+}|^{2}+(\sigma_{N}-1)|\Gamma_{+}|^{4}}
{|1+(\sigma_{N}-1)\Gamma_{+}^{\quad2}|^{2}} 
 \hspace{36pt}\mbox{$x$-$y$ plane interface}
\label{eqn:e8}
\end{eqnarray}
\[ \displaystyle
\Gamma_{\pm}=\frac{\Delta_{0}\sin\theta(\sin\phi\pm\cos\phi)}
{E+\Omega_{\pm}}, \hspace{12pt} 
\Omega_{\pm}=\sqrt{E^{2}-\Delta_{0}^{2}\sin^{2}\theta
(\sin\phi\pm\cos\phi)^{2}}.\]
\end{itemize}
\begin{itemize}
\item $E_{u}$(2)
\begin{eqnarray}
\sigma_{S,\uparrow}
\displaystyle
&=&\frac{1+\sigma_{N}|\Gamma|^{2}+(\sigma_{N}-1)|\Gamma|^{4}}
{|1-e^{-2i\phi}(\sigma_{N}-1)\Gamma^{2}|^{2}} 
 \hspace{12pt} \mbox{$z$-$y$ plane interface}\\[12pt]
\displaystyle
&=&\frac{1+\sigma_{N}|\Gamma|^{2}+(\sigma_{N}-1)|\Gamma|^{4}}
{|1+(\sigma_{N}-1)\Gamma^{2}|^{2}} 
 \hspace{12pt} \mbox{$x$-$y$ plane interface}
\end{eqnarray}
\[ \displaystyle
\Gamma=\frac{E-\Omega}{|\Delta_{0}\sin\theta|},  
\hspace{24pt}\Omega=\sqrt{E^{2}-\Delta_{0}^{2}\sin^{2}\theta}.\]
\end{itemize}
While $\sigma_{S,\downarrow}$ 
is unity due to the absence of the effective 
pair potentials. This feature is peculiar to the 
non-unitary superconducting state.
Figures 2 and 3 show the calculated conductance spectra of the two states
for various barrier heights.
Here, we assume that 
the injected electrons have equal probability weight for both 
up and down spin components, and $\bar{\delta}$ is chosen as 0.05$\pi$
to express the two-dimensional 
features of the Fermi surface.
In Fig. 2A, the magnitude of zero-bias conductance peaks 
increases with the increase of $Z$ as in our previous works. 
The origin of the zero-bias conductance peaks is that 
the denominator of the conductance formula in 
Eq.(\ref{eqn:e7}) vanish in the large $Z$ limit 
for $-\pi/4<\phi<\pi/4$. 
The zero-bias conductance peaks are universal properties 
for the junction of anisotropic superconductors 
independent of their parity and unitarity, 
where the pair potentials  change sign on the  
Fermi surface. 
On the other hand, for $x$-$y$ plane interface junction, 
the zero-bias conductance peaks do not appear(Fig. 2B). 
With the increase of $Z$, 
$\sigma(0)$ converges not to 0, but 0.5 
due to the residual density of states on the Fermi surface of 
quasiparticles with down spins. 
In the limit of two-dimensional Fermi surface, 
$i.e.$, $\bar{\delta} \rightarrow 0$, 
\begin{equation}
\sigma_{S,\uparrow}=\rm{Re}
\left[\frac{E}{\sqrt{E^{2} -\Delta_{0}^{2}[\cos\phi + \sin\phi]^{2}}}\right], 
\ \ 
\sigma(E) =\frac12\left[1 +  
\frac{1}{2\pi} \int^{2\pi}_{0} \sigma_{S,\uparrow} d\phi\right]
\label{eqn:e10}
\end{equation}
is satisfied. The obtained 
$\sigma(E)$ expresses the bulk density of states 
of $E_{u}(1)$ state superconductor. 
In the case 
of $E_{u}$(2) state with the $z$-$y$ plane interface, 
$\sigma(E)$ becomes maximum at $E=0$. 
The quantity $\sigma(0)$ increases with the increase of $Z$. 
In the limit of the large magnitude of $Z$, 
$\sigma(0)$ converges to a certain value which is larger than 
0.5 (Fig. 3A). In this case, 
the denominator of $\sigma_{S,\uparrow}$ vanishes  
at $E=0$  only for  $\phi=0$. 
Hence, the strong enhancement of 
$\sigma(0)$ with the increase $Z$ does not occur 
as in Fig. 2A.  
When the interface is perpendicular to the $z$-axis, 
the conductance spectra have a U-shaped structure (Fig. 3B) for 
larger $Z$. In this case, with the decrease of $\bar{\delta}$, 
$\sigma(E)$ converges to 
the bulk DOS of the $E_{u}(2)$ superconductors as in Fig. 2B. 

In this paper, we have studied the properties of
tunneling spectra in $N/I/TS$ junctions. 
Although our formula can be extended for any triplet superconducting states, 
the present paper mentions only about the results for pairing states which
are the most promising for Sr$_{2}$RuO$_{4}$. 
The existence of the large residual density of states of quasiparticles 
reflects the non-unitary superconducting states. 
The zero-bias conductance peaks and gap structures are obtained depending 
on the tunneling direction. 
By polarizing the injected electron with, for example,
a ferromagnetic normal metal, 
we can selectively measure the conductance 
spectrum components for the corresponding spin directions.
If the flat metallic spectra for the down spin injection
and  gap structures (or zero-bias conductance peaks) for 
the up spin injection are detected,
they can be regarded as the most clear evidence for the realization of
the non-unitary superconducting states.
We hope our theory will give a guide to determine the symmetry of 
the pair potential in Sr$_{2}$RuO$_{4}$.

\vspace{0.4cm}
One of the authors (Y.T.) is supported by a Grant-in-Aid for Scientific 
Research in Priority Areas 
''Anomalous metallic state near the Mott transition.'' 
The computational aspect of this work has been done for the 
facilities of the Supercomputer Center, Institute for Solid State Physics, 
University of Tokyo and the Computer Center, Institute for Molecular 
Science, Okazaki National Research Institute.

\begin{figure}
\begin{center}
\fbox{\epsfile{file=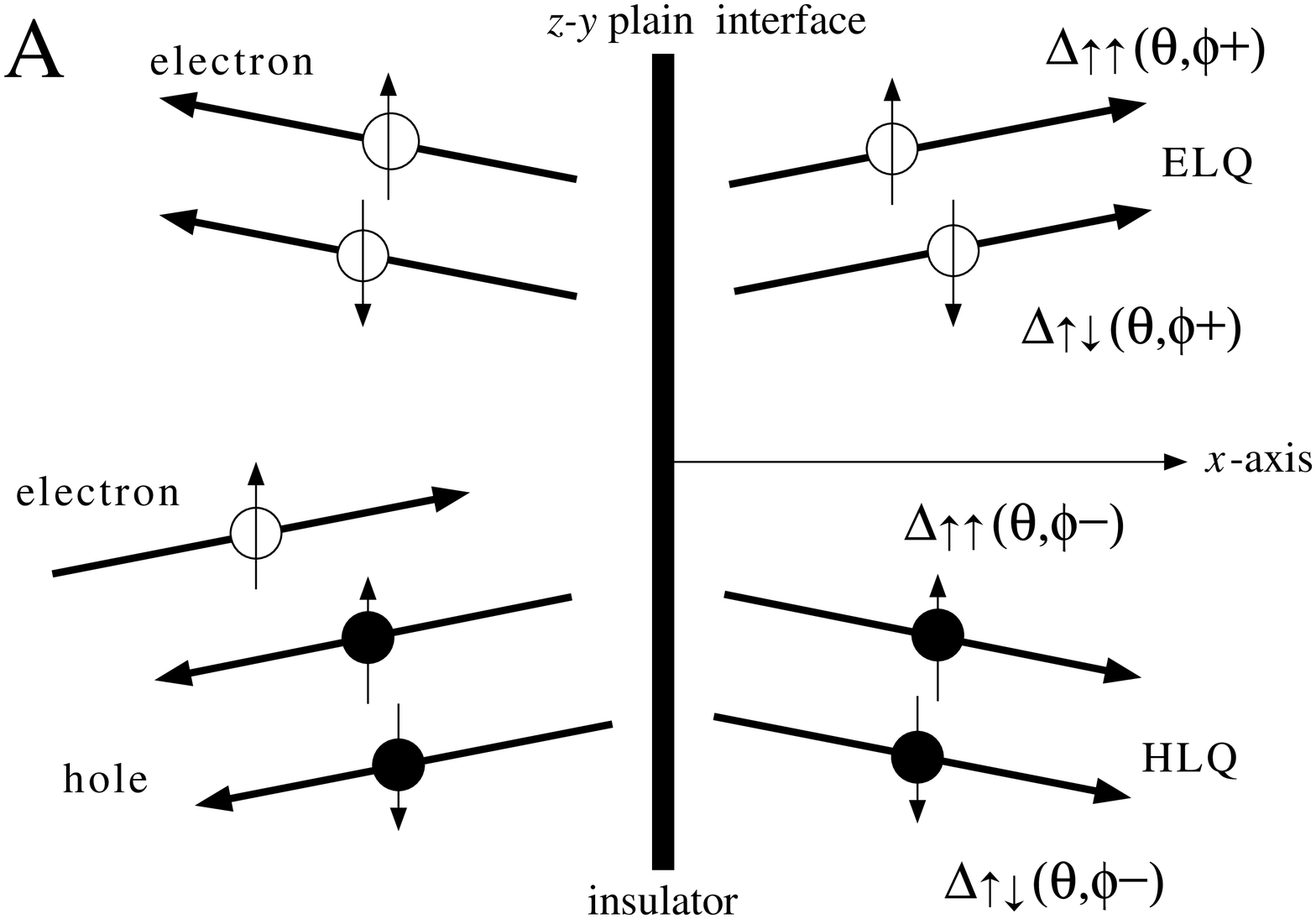,height=6cm,width=9cm,scale=1}}
\fbox{\epsfile{file=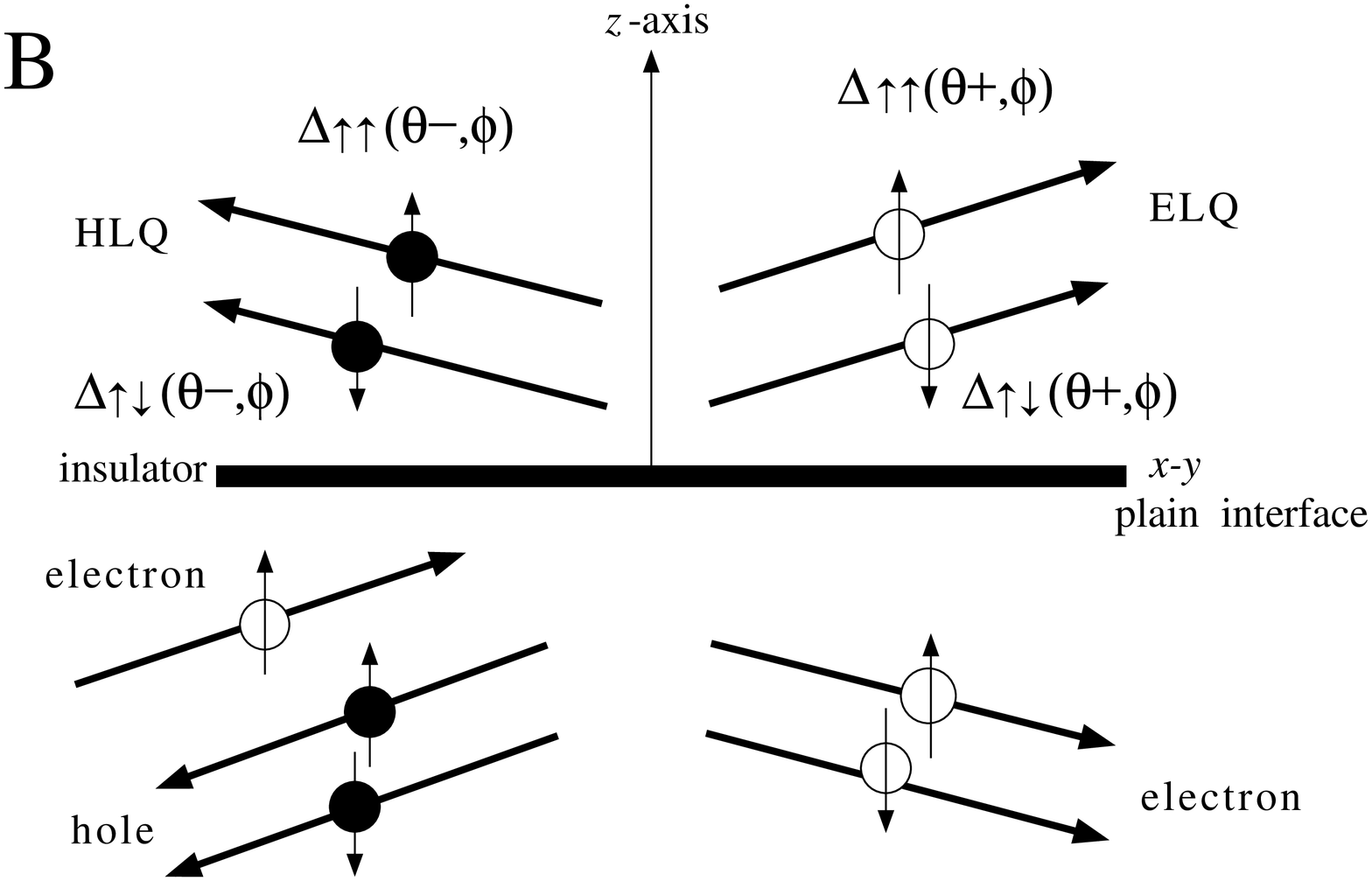,height=6cm,width=9cm,scale=1}}
\end{center}
\caption{Schematic illustration of the reflection and the transmission 
process of the quasiparticle at the interface of the 
junction with $z$-$y$ plane interface (Fig.1A)
and $x$-$y$ plane interface (Fig.1B). 
The $\theta$ and $\phi$ are the polar angle and azimuthal angle, 
respectively.}
\label{fig:f1}
\end{figure}

\begin{figure}
\begin{center}
\epsfile{file=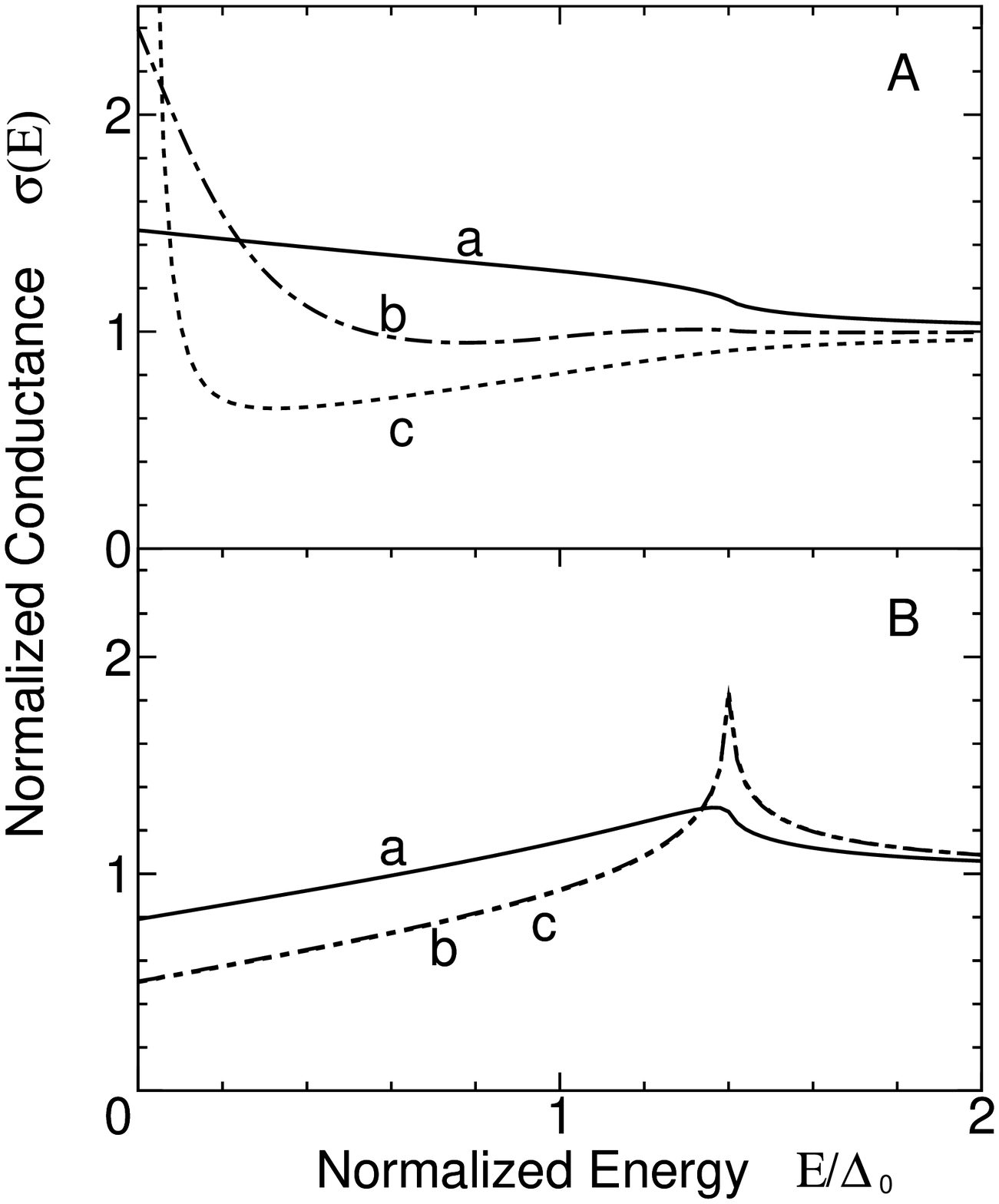,height=8cm,width=5cm,scale=1}
\end{center}
\caption{Normalized tunneling conductance is plotted for 
$E_{u}(1)$  state. 
A: $x$-axis is perpendicular to the interface 
($z$-$y$ plane interface). 
B: $z$-axis is perpendicular to the interface 
($x$-$y$ plane interface). 
a: $Z$=0.1, b: $Z$=1, and c: $Z$=5. }
\label{fig:f2}
\end{figure}

\begin{figure}
\begin{center}
\epsfile{file=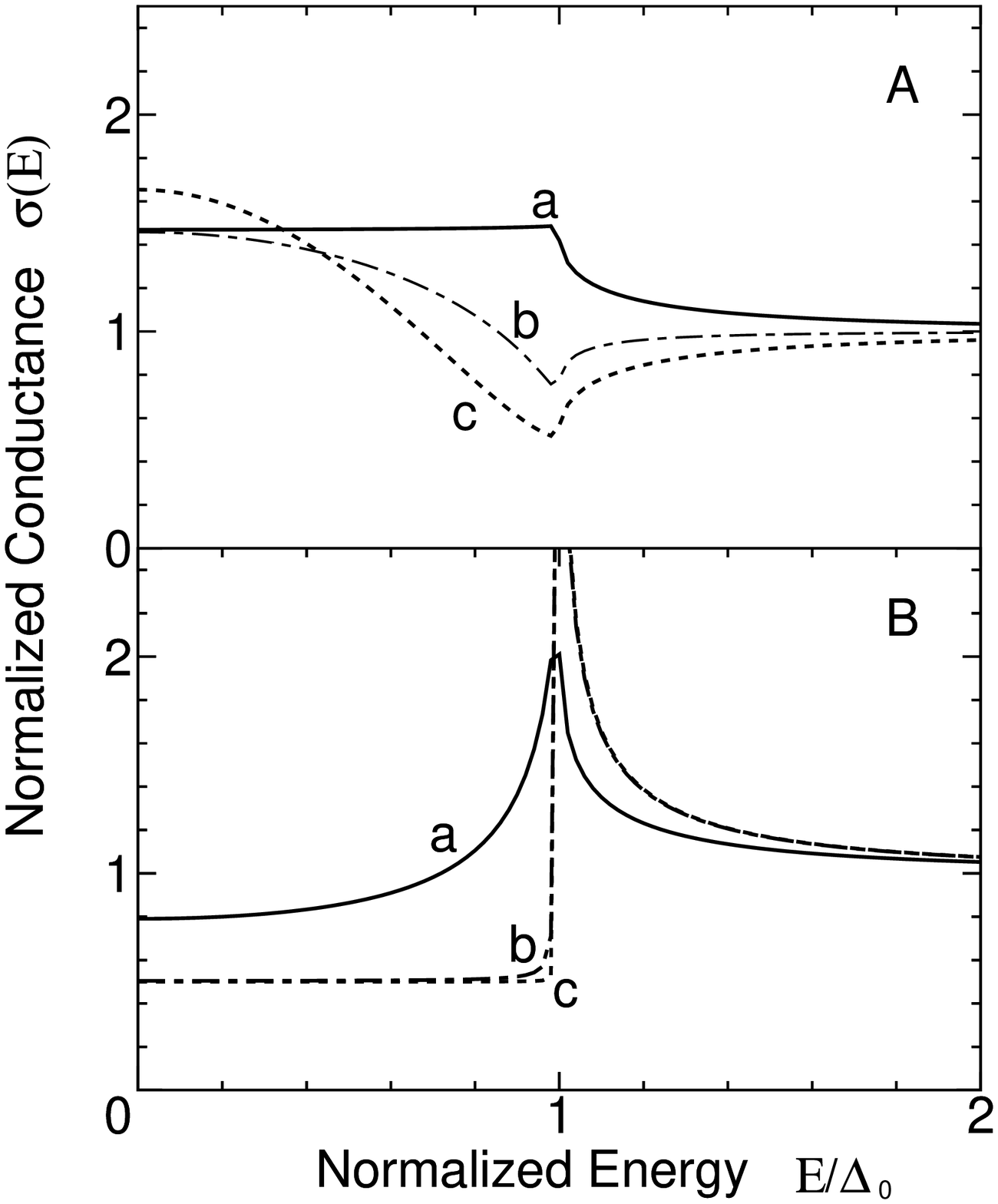,height=8cm,width=5cm,scale=1}
\end{center}
\caption{Normalized tunneling conductance is plotted for 
$E_{u}(2)$  state. 
A: $x$-axis is perpendicular to the interface 
($z$-$y$ plane interface).  
B: $z$-axis is perpendicular to the interface 
($x$-$y$ plane interface). 
a: $Z$=0.1, b: $Z$=1, and c: $Z$=5.}
\label{fig:f3}
\end{figure}

\end{document}